\documentclass [11pt]{article}
\usepackage{amsmath,amsthm,amsfonts,amscd,eucal,latexsym,amssymb}
\usepackage{epsfig}  
\def\cH{{\ca H}}

\def\cS{{\ca S}}

\def\fH{{\bf H}}

\def\cc{\text{c.c.}}

\def\bC{{\mathbb C}}           

\def\bM{{\mathbb M}} 
 
\def\bR{{\mathbb R}}

\def\bS{{\mathbb S}}

\newsymbol\rest 1316         
 
\def\gF{{\mathfrak F}}

\def\beq{\begin{eqnarray}}
\def\eeq{\end{eqnarray}}
\def\pa{\partial}
\def\at{\left(}               

\def\ct{\right)}              
\newcommand{\ca}[1]{{\cal #1}}         

\def\be{\beta}
\def\ga{\gamma}
\def\de{\delta}

\def\la{\lambda}

\def\om{\omega}

\def\La{\Lambda}
\def\Si{\Sigma}
\def\Om{\Omega}

\newcommand{\bra}[1]{\langle{#1}|}
\newcommand{\ket}[1]{|{#1}\rangle}
\newcommand{\nref}[1]{(\ref{#1})}
\newcommand{\media}[1]{\langle{#1}\rangle}

\newcounter{proposition}[section]
\newcounter{theorem}[section]
\newcounter{lemma}[section]
\newcounter{definition}[section]
\def\theproposition{\thesection.\arabic{proposition}}
\def\thetheorem{\thesection.\arabic{theorem}}
\def\thelemma{\thesection.\arabic{lemma}}
\def\thedefinition{\thesection.\arabic{definition}}

\newcommand{\se}[1]{\section{#1}}

\newcommand{\sse}[1]{\subsection{#1}}

\def\ssb #1 {\ifhmode{\par}\fi\refstepcounter{subsection}
  \noindent {\bf\thesubsection.} {\bf #1.}\quad
  \addcontentsline{toc}{subsection}{\protect\numberline{\thesubsection} #1}%
  }

\def\proposizione {\ifhmode{\par}\fi\refstepcounter{proposition}
  \noindent {\bf Proposition \theproposition}. \quad}
\def\teorema {\ifhmode{\par}\fi\refstepcounter{theorem}
  \noindent {\bf Theorem \thetheorem}. \quad}
\def\lemma {\ifhmode{\par}\fi\refstepcounter{lemma}
  \noindent {\bf Lemma \thelemma}. \quad}
\def\definizione {\ifhmode{\par}\fi\refstepcounter{definition}
  \noindent {\bf Definition \thedefinition}. \quad}

\begin{document} 
 
\hfill{\sl Preprint UTM 681 - May 2005} 
\par 
\bigskip 
\par 
\rm 
 
 
\par 
\bigskip 
\LARGE 
\noindent 
{\bf De Sitter quantum scalar field and horizon holography} 
\bigskip 
\par 
\rm 
\normalsize 
 
 
\large 
\noindent {\bf Nicola Pinamonti} \footnote{E-mail: pinamont@science.unitn.it} \\
 Department of Mathematics, Faculty of Science,
University of Trento, \&
 Istituto Nazionale di Alta Matematica ``F.Severi'',  unit\`a locale  di Trento \& Istituto
 Nazionale di Fisica Nucleare,  Gruppo Collegato di Trento, 
 via Sommarive 14, 
I-38050 Povo (TN), 
Italy\\\par 
\rm\large\large 
 
\rm\normalsize 

\rm\normalsize 
 
 
\par 
\bigskip 

\noindent 
\small 
{\bf Abstract}. 
We show explicitly that free quantum field theory in de Sitter background restricted on the cosmological horizon produces another quantum field theory unitarily equivalent with the original one. 
Symmetry properties descending from the dual theory are also remarked.
In the restricted theory the thermal properties, known for de Sitter quantum field theory, can be proved straightforwardly.

\normalsize
\bigskip 

\se{Introduction}
Motivated by the work of Guido and Longo \cite{gl2003}, we want to show the relation between the quantum field theory in de Sitter spacetime and another dual theory defined on its horizon.
Here the horizon is the cosmological horizon of a four dimensional de Sitter spacetime.
The interest in de Sitter spacetime comes in principle from cosmological consideration, in fact a small but not zero cosmological constant was recently measured.
Then it seems that the present form of the universe is compatible with a slice of a de Sitter universe.
On the other hand interest in the de Sitter spacetime arises in the framework of holographic theories.
The holographic principle was proposed by Susskind \cite{suss95} and argued by 
't Hooft \cite{thoo93, thoo95} for theories concerning black holes. It says that gravity can be described in a low dimensional quantum theory.
In that contest, the conjecture of Maldacena \cite{mald98}, further developed by Witten \cite{witt98},  arose. 
The conjecture is about the correspondence of Anti de Sitter spacetime and conformal field theories ($AdS/CFT$), in the framework of string theories. 
Some years later Rehren showed that the $AdS/CFT$ can be proved also for local quantum field theories \cite{rehren}.
A holographic description of de Sitter spacetime was searched.
Such a correspondence was proposed semiclassically by Strominger \cite{strominger}\footnote{Prviously, some work leading to dS/CFT was done by Mu-In Park in \cite{park1,park2}.}, then it was further developed by Klemm \cite{klemm}, Nojiri and Odintsov \cite{nood}.
The correspondence is between a bulk theory and theory located at the past infinity\footnote{See the work \cite{kv2002} and reference therein  for details. Another point of view regarding gravitational holography can be found in \cite{pad1,pad2}.}. Moreover, by this procedure, the theory of gravity is related to a Liouville theory located at past infinity. 
Then, the boundary theory can be quantized in a simpler way w.r. to quantization of gravity.
In this paper we show that bulk free massive quantum field theory in de Sitter universe can be mapped holographically on quantum field theory defined on the cosmological horizon. The mapping is implemented as a unitary transformation between Fock spaces. Moreover it has a geometrical meaning, in fact it correspond to the restriction of the bulk field on the horizon.
The quantum field theory in de Sitter spacetime was a longway subject of study \cite{BD} \cite{allen} \cite{allenfolacci} \cite{mottola} 
also considering only partial patches of de Sitter spacetime \cite{polarski}.
From a more abstract point of view we remind the works \cite{bmg,brosmoschella} in which generalized abstract field on the de Sitter spacetime are considered.
Another interesting correspondence between de Sitter generalized field and Minkowski fields was studied in \cite{bgms}.
The quantization of dS scalar field in relation with holography and conformal field theory, was also studied in \cite{gui1,lowe,gui2}.
In the next section we discuss some general facts about the four-dimensional de Sitter spacetime, everything is well known, but necessary to fix  notations we use in the paper.
Then we remind massive scalar quantum field theory on the de Sitter background.
In section 4 we pose the theory on the horizon and we study its relation with the theory in the de Sitter spacetime some comments on symmetries are remarked through holography.
Finally, by means of holography, some commets about the thermal properties are discussed.

\se{de Sitter spacetime}
In this section we recall some well known results about four dimensional de Sitter spacetime\footnote{For a comprehensive review see \cite{stro2001}, or \cite{kv2004}}.
We restrict our discussion to the four dimensional case but some facts we are presenting here are also valid in the $d-$dimensional case. 
De Sitter (dS) spacetime is the solution of vacuum Einstein equations with positive cosmological constant $\La=\frac{6}{\ell^2}$ where $\ell$ has the dimension of a length.
It can be seen as an hyperboloid immersed in a 5-dimensional Minkowskian spacetime $\bM^5$.
Assuming that the natural metric $\ga$ of $\bM^5$ has signature $(-,+,+,+,+)$, 
de Sitter spacetime is defined as the set of Minkowskian vectors $X$ satisfying: 
$$
X^a X^b\ga_{ab} = \ell^2.
$$
The metric induced on the hyperboloid defines the de Sitter metric.
Cartan Penrose conformal diagram of the dS spacetime is plotted in the Figure above, where horizontal lines correspond to 3-d sphere, whose
extremal points are 2-d spheres (the south and the north pole).
\begin{figure}
\begin{center}
\epsfig{file=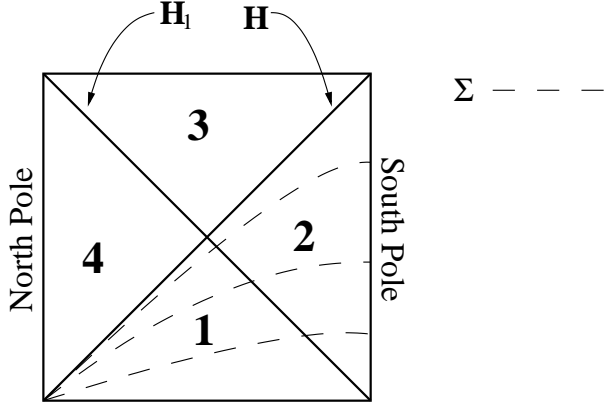,width=.5\textwidth}
\end{center}
\caption{Figure: Cartan Penrose conformal diagram of de Sitter spacetime}
\end{figure}
Notice from the Figure that dS spacetime posses the cosmological horizons $\fH$ and $\fH_1$, in fact
an observer situated in the north pole cannot send information to the south pole in a finite time, as it appears clear from the conformal diagram.
There is a global coordinate system, but it is not static. Moreover the quantization procedure seems more difficult in that coordinate system, for this reason, in the next we shall deal with two patch of the de Sitter metric. First of all consider the following parametrization 
of dS spacetime called the {\it planar coordinate} system.
$$
X^0=\ell\sinh{\at\frac{\tau}{\ell}\ct}-\frac{e^{-\frac{\tau}{\ell}}}{2\ell} R^2,\quad 
X^i=e^{-\tau} R \om^i,\quad 
X^4=\ell\cosh{\at\frac{\tau}{\ell}\ct} -\frac{e^{-\frac{\tau}{\ell}}}{2\ell} R^2,
$$
where $i\in [1,2,3]$ and $\om^i$ are the standard coordinate\footnote{$\om^1=\cos(\theta)$, $\om^2=\sin(\theta)\cos(\varphi)$, $\om^3=\sin(\theta)\sin(\varphi)$ } for a two dimensional unit sphere immersed in $\bR^3$.
The line element in this coordinate system has the following form:
\beq\label{planar}
ds^2=-d\tau^2+e^{-2\frac{\tau}{\ell}}\at dR^2+R^2d\Om^2\ct,
\eeq
where $d\Om$ is the standard metric  of the 2-dimensional sphere, $\tau\in(-\infty,\infty)$ and $R>0$.
This patch covers only half of de Sitter spacetime, precisely the regions 1 and 2 of the Figure. We can extend conformally this metric considering the following transformation $\eta=\ell e^{\tau/\ell}$ with $\eta=(0,+\infty)$. 
De Sitter metric becomes conformally equivalent to the four dimensional Minkowskian metric.
$$
ds^2= \frac{\ell^2}{\eta^2}\at-d\eta^2+dR^2+R^2d\Om^2\ct.
$$
$\eta=0$ correspond to past infinity while $\eta\to \infty$ identifies only the the south pole of future infinity (see Figure).
The cosmological horizon $\fH$ is the null type boundary of that patch.
It is reached when $\eta$ and $R$ tend to infinity while $\eta-R=V$ is finite.
As emerges from the line elements both the coordinate system introduced before are not static. 
Considering $r=\ell R/\eta$, assumed positive, and $t=\frac{\ell}{2} \log{\at\frac{\eta^2-R^2}{\ell^2}\ct}$, 
we get a {\it static coordinate system}, in fact the line element \nref{planar} reads in the new coordinate:
\beq\label{static}
ds^2=-\at 1-\frac{r^2}{\ell^2} \ct dt^2+\at 1-\frac{r^2}{\ell^2} \ct^{-1} dr^2+r^2 d\Om^2.
\eeq
This patch covers only the region $2$ of the Figure. It is bounded by the south pole and by half of the future $\fH$ and past $\fH_1$ cosmological horizons.
Since the metric does not depend on $t$, 
$$
\frac{\pa}{\pa t}= \frac{\eta}{\ell}\frac{\pa}{\pa \eta}+\frac{R}{\ell}\frac{\pa}{\pa R}
$$
is  a Killing vector. It can be extended in the whole dS spacetime however is time like in the region 2 and 4, spacelike in region 1 and 3. It is  lightlike on $\fH$ and $\fH_1$. For this reason it does not define a global Killing time.
It is important anyway because it describes a static observer. Moreover it is the unique (up to constant factor) Killing vector 
of the form $A(\eta,R)\frac{\pa}{\pa \eta}+B(\eta,R)\frac{\pa}{\pa R}$.

\se{Free massive quantum scalar field on ds Sitter spacetime}
In this section we present canonical quantization of scalar fields minimally coupled with the metric.
We perform computations in planar coordinates \nref{planar}.
It wants to be mainly an introduction in the notation used below. In fact quantization of scalar fields on de Sitter spacetime is well known\footnote{See for example \cite{BD,allen,mottola,allenfolacci,polarski} and reference therein.}, also from an abstract algebraic point of view \cite{bmg,brosmoschella}.

\sse{One-particle Hilbert space.}
The first step in the quantization procedure is the definition of the one-particle Hilbert space.
To this end we consider Klein Gordon equation in planar coordinates:
\beq\label{KGplan}
-\frac{\eta^4}{\ell^2}\frac{\pa}{\pa \eta} \frac{1}{\eta^2}\frac{\pa}{\pa \eta} \psi + \frac{\eta^2}{\ell^2}\at \frac{1}{R^2}\frac{\pa}{\pa R} R^2\frac{\pa}{\pa R} \psi -\frac{L^2}{R^2} \psi   \ct  =\la^2\psi,
\eeq
where $L^2$ is the usual angular momentum operator.
$\la$ encompass both the mass of the field $M$ and the coupling with the curvature $\xi$: $\la^2=M^2+12\,\xi/\ell^2$.
Real compactly supported smooth solutions $\psi$  of the equation \nref{KGplan} represent the {\bf wavefunctions} associated with particles.
To build the one-particle Hilbert space $\cH$ we consider the frequency decomposition of the real scalar fields $\psi$ w.r. to the time $\eta$, then
$\cH$ is defined as the completion of the $\eta$-positive frequency part $\psi_+$ of $\psi$ w.r. to 
the scalar product
\beq\label{scalar}
\langle\psi_+,\psi'_+\rangle:=
i\int_\Sigma \at\: \overline{\psi_+}\: \nabla_\mu \psi'_+- {\psi'_+}\: \nabla_\mu \overline{\psi_+}\ct n^\mu \sqrt{\sigma} d\sigma 
=i\int_\Sigma \at\: \overline{\psi_+}\: \pa_\eta \psi'_+- {\psi'_+}\: \pa_\eta \overline{\psi_+} \ct  \at\frac{R \ell}{\eta}\ct^2 dR\:d\Om.
\eeq
Above $\Sigma$ is a $\eta$-constant hypersurface in de Sitter spacetime, $d\sigma$ is the volume form descended from the metric induced on $\Sigma$ and  $n$ is the unit vector orthogonal to $\Sigma$ pointing toward the future. It is possible to show that the procedure is independent on the choice $\eta$ characterizing $\Sigma$. Notice that $\Si$ is a Cauchy surface for the submanifold of the dS spacetime composed by the part 1 and 2 of the Figure. 
Referring to the procedure described above the mode decomposition of the real solution of \nref{KGplan} reads:
\beq\label{modes}
\psi(\tau,R,\theta,\varphi)=
\int_0^{+\infty} \sum_{lm}   
Y^m_l(\theta,\varphi)
\Phi_{l\nu}(E,R,\eta) 
\widetilde{\psi^+_{lm}}(E) 
+
Y^m_l(\theta,\varphi)
\overline{\Phi_{l\nu}(E,R,\eta) }
\overline{\widetilde{\psi^+_{lm}}(E)}\;
dE.
\eeq
where $Y_l^m(\theta,\varphi)$ are the standard spherical harmonics normalized to one.
From now on $\sum_{lm}$ means $\sum_{l\geq0,m\in[-l,l]}$.
Moreover we have used the set of orthonormal modes:
\beq\label{mm}
\Phi_{l\nu}(E,R,\eta):= \sqrt{\frac{\pi E}{ R}} 
\frac{\eta^{\frac{3}{2}}}{2\ell}
H^{(2)}_{\nu} \at {E} \eta  \ct
J_{l+\frac{1}{2}} \at {E} R  \ct, \qquad 
\overline{\Phi_{l\nu}(E,R,\eta)}:= \sqrt{\frac{\pi E}{ R}} 
\frac{\eta^{\frac{3}{2}}}{2\ell}
H^{(1)}_{\nu} \at {E} \eta  \ct
J_{l+\frac{1}{2}} \at {E} R  \ct
\eeq
$J_\mu(x)$ is the Bessel function of the first kind and $H^{(\cdot)}_{\mu}(x)$ the Bessel functions of the third kind also known as Henkel functions\footnote{$H^{(1)}_{\mu}:=J_\mu(x)+iN_\mu(x)$, $H^{(2)}_{\mu}:=J_\mu(x)-iN_\mu(x)$.}, 
$\nu=\sqrt{\frac{9}{4}+\ell^2\;\la^2}$.
In $R=0$ there is a spatial coordinate singularity, then we consider only modes such that $\phi(\eta,0,\theta,\varphi)=0$,  once the space, spanned by these modes, is completed through the Hilbert completion, the whole Hilbert space is recovered.
Furthermore, when $E,\eta\to \infty$, the behavior  $H^{(2)}_{\nu} \at {E} \eta  \ct\sim e^{-iE\eta}/\sqrt{E\eta}e^{i\alpha}$ is in agreement with the usual adiabatic prescription \cite{BD}. The part of $\psi$ in \nref{modes} containing $H^{(2)}_\nu$ acts as positive frequency mode, whereas the one containing $H^{(1)}_{\nu} \at {E} \eta  \ct\sim e^{+iE\eta}/\sqrt{E\eta}e^{-i\alpha}$ acts as negative frequency part (here $\alpha$ is a phase we shall compute later, it does not enter in our considerations). 
Moreover the scalar product \nref{scalar} do not mix negative and positive frequencies, that is :
$\langle\psi_+,\psi_-\rangle=0$.
Furthermore {\it a posteriori} the scalar product \nref{scalar} is independent on the time $\eta$, writing
$$
\psi^R_+(R,\theta,\varphi)= \sum_{lm}\; \int_0^\infty E  J_{l+\frac{1}{2}}({E} R) {\widetilde{\psi^+_{lm}}(E)}\; Y^m_l(\theta,\varphi)\;dE   
$$ 
the scalar product on $\Si$ can be written as: 
\beq
\langle \psi_+,\psi'_+\rangle:= \int_\Si \overline{\psi^R_+(R,\theta,\varphi)} {\psi^R_+}'(R,\theta,\varphi) \; R\;dR\; d\Om
\eeq
non involving $\eta$ at all. Then $H^{(1)}$, $H^{(2)}$ play the role of the exponentials $e^{\pm iEt}$ in the usual mode decomposition of real field in the Minkowski spacetime.
Notice that the modes \nref{modes} are orthogonal and complete. 
We can read the quantum theory in $E$ variable. In terms of the positive frequency part $\widetilde{\psi^+_{lm}}(E)$ the scalar product \nref{scalar} becomes
$$
\langle \psi_+ ,\psi'_+ \rangle=\sum_{lm}\int_0^{+\infty} \overline{\widetilde{\psi^+_{lm}}(E)} \widetilde{\psi^+_{lm}}'(E)\; dE.
$$
A straightforward analysis proves that the one-particle Hilbert space is $\bigoplus_{l=0}^\infty \at L^2(\bR^+,dE)\otimes\bC^{2l+1}\ct$.
$\bC^l$ represent the degeneracy induced by the angular quantum number $m$.\\

\sse{Fock space.} 
As usual from the one particle Hilbert space $\cH$ we build up the bosonic Fock space $\gF_+(\cH)$.
The bosonic quantum field, acting on $\gF_+(\cH)$, is defined, in terms of \nref{mm}, as following:
\beq\label{field}
\hat{\phi}(\eta,R,\theta,\varphi)= \int_{0}^{+\infty}\:\sum_{lm}\:     \Phi_{l\nu}(E,R,\eta)\: Y^m_l(\theta,\phi)\:   a_{Elm} + \overline{\Phi_{l\nu}(E,R,\eta)} 
\: \overline{Y^m_l(\theta,\phi)}\:
a^\dagger_{Elm}\: dE.
\eeq
where $a_{Elm},a_{Elm}^\dagger $ are respectively the annihilation and creation operators satisfying the standard commutation relation
$[a_{Elm},a_{E'l'm'}^\dagger]:=\delta(E-E') \de_{ll'}\de_{mm'}$.
The ground state $\ket{0}$ of the theory is that it satisfy 
$$
a_{Elm}\ket{0}=0
$$
for every quantum numbers $E,l,m$.\\
\noindent {\bf Remarks:}\\
\noindent {\bf (a)}
The symmetrized two point function \cite{BD,allenfolacci} is
$$
G_+(x,x')=\bra{0} \hat{\phi}(x) \hat{\phi}(x') +\hat{\phi}(x') \hat{\phi}(x)  \ket{0}=\frac{\at\frac{1}{4}-\nu^2\ct}{16\pi\ell^2}\frac{1}{\cos(\pi\nu)} 
F\at\frac{3}{2}+\nu,\frac{3}{2}-\nu;2;\frac{1+Z(x,x')}{2}\ct
$$
$F$ is the hypergeometric function, and $Z(x,x')$ is the geodesic distance between $x$ and $x'$. Since
the two-point function depends on the geodesic distance between the points $x$ and $x'$ only, the ground state $\ket{0}$ is invariant 
under $SO(1,4)$ transformations (the dS group).

\noindent {\bf (b)}
We remark that our choice of vacuum, depending on the modes decomposition \nref{modes}, is {\it a posteriori} equivalent to the choice of the natural vacuum in global coordinates \cite{BD}.

\noindent {\bf (c)} The asymptotic behavior of $\Phi_{l\nu}(E,R,\eta)$ \nref{mm} near the cosmological horizon $\fH$ ($\eta+R \to \infty$ keeping fixed $\eta-R=V$ ) is
\beq\label{asym}
\Phi_{l\nu}(E,R,\eta)\sim \at\frac{\eta}{R\ell}\ct\;\frac{e^{-iE(\eta-R)}}{\sqrt{4\pi E}} e^{i\frac{\pi}{2}\at\nu-l-\frac{1}{2}\ct}\:.
\eeq
This suggests that restricting the field on the horizon $\fH$ is possible to get an equivalent quantum theory holographically related with that presented above. This is the main topic of this paper and it will be analyzed below.

\noindent {\bf (d)} Of course, to have a well defined field operator, the field \nref{field} needs to be smeared.
The most simple way to implement field smearing is in terms of wavefunction. Then considering a real wavefunction (defining the particle) $\psi$
the field smeared with $\psi$ reads:
$$
_w\hat{\phi}(\psi):=\int_\Sigma \hat{\phi}(\eta, R ,\theta,\varphi)\:\frac{\pa}{\pa\eta}\:  \psi(\eta,R,\theta,\varphi)- 
\frac{\pa}{\pa\eta}\: \hat{\phi}(\eta, R ,\theta,\varphi)\: \psi(\eta,R,\theta,\varphi) \:
 dR\; d\Om\;.
$$
To implement locality we have to introduce the unique causal propagator $E$ in the globally hyperbolic region we are considering satisfying 
the usual properties\footnote{For a more comprehensive description of that procedure see \cite{wald}.}: $E$ maps real compactly supported functions $f$ to real wavefunctions $\psi$ in a linear and surjective way, $\text{supp}(Ef) \subset J(\text{supp} f)$, $Ef=0$ if and only if $f=Kg$ where $K$
is the Klein Gordon operator\footnote{$K\psi:=\Box\psi-\la^2\psi$.}.
Then field smeared by function $f$ is defined as $\hat{\phi}(f):=_w\hat{\phi}(Ef)$.
It follows that 
$$
\hat{\phi}(f):=\int_{dS} f(\eta, R ,\theta,\varphi) \hat{\phi}(\eta, R ,\theta,\varphi)\; \sqrt{g}\; d\eta\; dR\; d\Om.
$$

\se{Horizon Quantum Field}
The aim of this section is to define the scalar quantum field theory on the horizon.
The idea of building a theory on null surfaces as horizon was presented in a very nice work of Sewell \cite{sewell}, 
it was further developed in \cite{MP2,MP3,MP4,MP6}, we follow here that ideas.\\
The horizon of a four dimensional dS spacetime is topologically a line times a two sphere ($\fH\sim\bR\times \bS^2$). $\bS^2$ can be equipped with the natural metric of the two-sphere whose line element is $d\Om$ whereas there is not possible to fix a natural metric on $\bR$. As a consequence, the scalar product needs to be defined carefully on it.

\sse{One-particle Hilbert space.}
A point in $\fH$ is described by the coordinates $(V,\theta,\varphi)$ where $(\theta,\varphi)$ are the usual coordinate of $\bS^2$ while $V$ is a generic coordinate in $\bR$, up to now it has no particular meaning. In fact because of the absence of the metric on $\bR$ there is no preferred (natural) coordinate system.
Consider the classical real scalar fields on the null surface $\fH$:
\beq\label{nulldec}
\psi_\fH(V,\theta,\varphi)= \int_{\bR^+} \frac{e^{-iEV}}{\sqrt{4\pi E}} \sum_{lm}  e^{i\rho_{lm}} Y^m_l(\theta,\varphi)  \widetilde{\psi^+_{\fH\,lm}}(E) + \cc.
\eeq
$\rho_{lm}$ represent a fixed phase depending on the angular momentum quantum numbers $l,m$. 
$\psi_\fH(V,\theta,\varphi)\in\cS(\fH)$, where $\cS(\fH)$ is the set of function on $\fH$ made as the product of a real valued Schwartz function of the line and a smooth function of the sphere $\bS^2$.
We shall use the positive part decomposition suited above as the definition of {\it horizon particle} forming the horizon Hilbert space $\cH_{\fH}$. 
In other more precise words, the Hilbert space $\cH_{\fH}$ is the completion of the space spanned by complex combination of wavefunctions  $\widetilde{\psi^+_{\fH\, lm}}$ descending from $\psi_\fH\in\cS(\fH)$, w.r. to the scalar product:
\beq\label{nullscalprod}
\langle\psi_{\fH+},\psi'_{\fH+}\rangle:=i\int_\bR \overline{\psi_{\fH+}(V,\theta,\varphi)} \pa_V \psi'_{\fH+}(V,\theta,\varphi) - \psi'_{\fH+}(V,\theta,\varphi)\pa_V\overline{\psi_{\fH+}(V,\theta,\varphi)}  dV d\Om.
\eeq
Notice that also in this case the ``$E$-modes $\sim e^{-iEV}$'' are orthogonal and complete, then the horizon one particle Hilbert space
$\cH_{\fH}$ is isomorphic to $\bigoplus_{l=0}^\infty \at L^2(\bR^+,dE)\otimes\bC^{2l+1}\ct$.
The scalar product \nref{nullscalprod} on $\cH_{\fH}$ reads 
$$
\langle\psi_{\fH+},\psi_{\fH+}'\rangle:=\int_{\bR^+} \sum_{lm} \overline{\widetilde{\psi^+_{\fH\,lm}}}\widetilde{{\psi^+_{\fH\,lm}}'} dE.
$$

\sse{QFT on the Fock space.}
In terms of the decomposition \nref{nulldec}, on the Fock space $\gF_+(\cH_\fH)$ equipped with the natural vacuum $\ket{0}_\fH$ the canonical quantum field reads:
\beq\label{horizonfield}
\hat{\phi}_\fH(V,\theta,\varphi)= \int_{\bR^+} \frac{e^{-iEV}}{\sqrt{4\pi E}} \sum_{lm}  e^{i\rho_{lm}} Y^m_l(\theta,\varphi)  
a^\fH_{Elm}+ \frac{e^{+iEV}}{\sqrt{4\pi E}} \sum_{lm}  e^{-i{\rho_{lm}}} \overline{Y^m_l(\theta,\varphi)}  
{a^\fH}^\dagger_{Elm}.
\eeq
where as above ${a^\fH}^\dagger_{Elm},{a^\fH}_{Elm}$ are the creator annihilator operators satisfying the canonical commutation relation $[{a^\fH}_{Elm},{a^\fH}^\dagger_{E'l'm'}]:=\de(E-E')\de_{ll'}\de_{mm'}$.\\

\noindent {\bf Remarks:}\\
\noindent {\bf (a)} As usuall to have a well-defined quantum field, it needs to be smeared. On the horizon the usual smearing is not defined because there is no a preferred measure (the metric being degenerate). A way to circumvent the problem is to implement a field smearing 
using forms instead of functions.
Then, consider a one form $\om(V,\theta,\varphi)=df(V,\theta,\varphi)$ on $\fH$, the smeared field is  
$$
\hat{\phi}_\fH(df):= \int\:\hat{\phi}_\fH(V,\theta,\varphi) \:df \wedge d\Om=
-\int f\: d \hat{\phi}_\fH(V,\theta,\varphi) \wedge   d\Om.
$$
Above in the last equality we have supposed that $f(V)$ vanishes at $\pm\infty$. It appears clear that the basic object of the theory is not the field itself. Because of the smearing, the quantum field theory is in fact built up by means of $d\hat{\phi}$ or, 
$\pa_V\hat{\phi}(V)$ once a coordinate $V$ is chosen. Notice that the same happens in conformal field theory.

\noindent {\bf (b)} Horizon quantum field on $\gF_+(\cH_\fH)$ enjoys symmetry under $SL(2,\bR)$ (conformal group). In fact, it is possible to prove that the Hilbert space $L^2(\bR^+,dE)$ admits a  unitary representation of $SL(2,\bR)$ generated by \cite{MP2,MP3} 
\beq\label{genop}
H:=E, D:=-i\at\frac{1}{2}+E\frac{d}{dE}\ct , -\frac{d}{dE}E\frac{d}{dE}+\frac{\at {k}-\frac{1}{2}\ct^2}{E}.
\eeq
Above ${k}$ is fixed in $\{1/2,1,3/2,\dots\}$, it labels a particular $SL(2,\bR)$ representation. The following commutation relations
of $sl(2,\bR)$ hold
$$
[H,D]=iH,\qquad [C,D]=-iC,\qquad [H,C]=2iD.
$$
Moreover the finite action of $H$, $D$ and $C$ on $\cH$ has a geometrical meaning in terms of particular diffeomorphisms transforming the $V$ coordinate: 
$$
e^{-iH\alpha}\psi(V,\theta,\varphi)= \psi(V+\alpha,\theta,\varphi), \qquad e^{-iD\alpha}\psi(V,\theta,\varphi)= \psi(e^{\alpha}V,\theta,\varphi),
$$
$$
e^{-iC\alpha}\psi(V,\theta,\varphi)=\psi \at \frac{V}{V+\alpha},\theta,\varphi\ct +A,
$$
$A$ is a constant, see \cite{MP2,MP3} for details. 
That action can be extended to the Fock space $\gF_+(\cH_\fH)$, then the constant $A$ on the action of $C$ can be discarded because of the form smearing. 
$SL(2,\bR)$ acts on the coordinate as the projective group.
Notice that even if the scalar product seems to be invariant under general reparametrization ($Diff^+(\cH_\fH)$ invariance) of the null coordinate $V$, the quantum field theory is not. The reason is that the decomposition in modes \nref{nulldec} depends (partially) on the choice of the coordinate $V$.
As a consequence $Diff^+(\cH_\fH)$ invariance cannot be realized as set of unitary transformations in the Fock space $\gF_+(\cH_\fH)$ \cite{MP4}.
\\
\noindent {\bf (c)} 
The causal propagator on $\fH$ reads
$$
[\hat{\phi}_\fH(V,\theta,\varphi),\hat{\phi}_\fH(V',\theta',\varphi')]:=-\frac{i}{4} sign(V-V') \de(\theta-\theta') \de(\varphi-\varphi').
$$
\noindent {\bf (d)} For completeness and for analyzing the thermal properties it is usefull to compute the two point function:
\beq\label{twopoint}
G_\fH:=\media{\pa\hat{\phi}_\fH(V,\theta,\varphi)\pa\hat{\phi}_\fH(V',\theta',\varphi')}_{\fH}:=-\frac{1}{4\pi} \frac{\de(\theta-\theta') \de(\varphi-\varphi')}{(V-V')^2}.
\eeq
It could appear strange to have a two point function written in term of $\pa\hat{\phi}$, but it is a consequence of
the adopted form-smearing procedure.
That is because of metric degenerateness on the horizon.

\se{Holography and conformal symmetry}
We are ready to present the main result of that paper. In fact we are to show that quantum theory of free massive scalar fields on de Sitter spacetime is in relation with the theory on the cosmological horizon presented above.
Moreover that relation has a deep geometrical and physical meaning. First of all the relation is described by a unitary operator mapping the theory in the bulk to the theory on the horizon. Then, most important, the action of the unitary operator on the field corresponds geometrically to a restriction of the dS field on the horizon.
To understand better both the claims consider the following unitary transformation $U$ mapping the dS Fock space $\gF_+(\cH)$ in the horizon Fock space
$\gF_+(\cH_\fH)$, with the following properties:
\beq
U\ket{0}=\ket{0}_\fH,\qquad U\: \hat{\phi}(\eta,R,\theta,\varphi)\: U^\dagger = \hat{\phi}_\fH(V,\theta,\varphi),
\eeq
where $V=\eta+R$. 
Furthermore $U$ written above fixes the phase $\rho_{lm}$ in \nref{horizonfield} to the value $\rho_{lm}:=\rho^\nu_{l}:=\frac{\pi}{2}\at\nu-l-\frac{1}{2}\ct$. Notice that the phase $\rho^\nu_{l}$ encompasses the information about the field mass
$M$ and the curvature coupling $\xi$, through $\nu$ . 
The action of $U$ on the creation and annihilation operator is very simple: $U a_{Elm} U^\dagger := a^\fH_{Elm}$. Moreover for the one-particle Hilbert spaces $(U\;\psi)(\eta,R,\theta,\varphi)=\psi_H(V,\theta,\varphi)$.
But there is another interesting geometrical property enjoyed by $U$: identifying $a^\fH_{Elm}$ with $a_{Elm}$ and remembering the asymptotic behavior \nref{asym} of the bulk modes \nref{mm}
$$
 U\: \hat{\phi}(\eta,R,\theta,\varphi)\: U^\dagger = \hat{\phi}_\fH(V,\theta,\varphi)=\left.\hat{\phi}(\eta,R,\theta,\varphi)\right|_\fH.
$$
Notice that we have realized a $dS/CFT$, indeed we have found that there is a unitary relation between bulk quantum field and the one on the horizon. 
Moreover, for the remark suited in the previous section the horizon theory is a invariant under the conformal group $SL(2,\bR)$.
The following facts merit particular remark.

\noindent {\bf Remarks:}\\
\noindent {\bf (a)} 
The unitary holographic relation presented here is a special case of the more general relation existing between the quantum observables algebra on the dS spacetime and on the horizon induced by $U^\dagger\hat{\phi}(f)U=\hat{\phi}_\fH(\om_f)$ where $\om_f=\left.2d((Ef)\right|_{\fH})$.

\noindent {\bf (b)} 
As a consequence, one has the holographic relation:
 the invariance of vacuum expectation values,
\beq
 _{\fH}\langle 0| \hat\phi_{\fH}(\om_{1})\cdots \hat\phi_{\fH}(\om_{n})|0\rangle_{\fH}
 = \langle 0| \hat\phi(f_1)\cdots \hat\phi({f_n})|0\rangle\:,
\eeq
where $\om_i=\left.2d((Ef_i)\right|_{\fH})$

\noindent {\bf (c)} 
As pointed out in the previous section 
a representation of $SL(2,\bR)$  acts unitarily
on $\gF_+(\cH_\fH)$, then, through $U$, a similar representation can be indiced on $\gF_+(\cH)$, that is that $SL(2,\bR)$ acts unitarily also on $\gF_+(\cH)$ the space of dS quantum fields.

\noindent {\bf (d)} 
Notice that the action of $D$, defined in \nref{genop}, on $\widetilde{\psi^+_{lm}}(k)$ correspond to the action of 
$$
D=i\eta \frac{\pa}{\pa \eta} + iR \frac{\pa}{\pa R}
$$
on $\psi_+(\eta,R,\theta,\varphi)\in\cH$ that correspond to the Killing vector field
$
d=\eta \frac{\pa}{\pa \eta} + R \frac{\pa}{\pa R}.
$
The other quantum operator $H,C$ have no meaning in terms of Killing vectors. In fact 
there is no Killing vector preserving spherical symmetry and non commuting (by means of Lie bracket) with $d$, as
$H$ and $C$ do with $D$.
Then the Lie algebra $sl(2,\bR)$ formed by $H,C,D$ \nref{genop} cannot be represented by any Lie subalgebra of the Killing vectors fields.
$H,C$ have no geometrical meaning in terms of Killing fields.
Finally notice that $d/\ell$ nothing but is $\pa/\pa t$ the generator  of translation of the Killing time $t$ then $D/\ell$ generates 
$t$-translations.

\noindent {\bf (e)} 
When $D$ can be considered as Hamiltonian of the system (for example considering the dS quantum field theory restricted in the region 2), $SL(2,\bR)$ represent an {\it hidden symmetry} for it in the sense that the transformation generated by $H,C$ have no geometrical meaning in terms of action of Killing vectors\footnote{See \cite{MP3} for a general discussion on symmetries.}. That symmetry becomes manifest considering the horizon theory through holography.

\se{An application: Thermal properties}
The holographic relation presented above is useful to simplify some computations. 
Here we show how the thermal properties of field theory become very simple on the horizon dual theory.
Since we deal with infinite volumes (thermodynamical limit), we say that a quantum theory is thermal
if its reference state satisfies the KMS (Kubo Martin Schwinger) condition.
In this section we shall show that the horizon quantum field theory dual to the dS one satisfy the KMS property with respect to the dilation at inverse temperature $2\pi\ell$.
An observer evolving under static Killing time $t$ measures an inverse temperature $\beta=2\pi\ell$.
In the one-particle Hilbert space $\cH$, that evolution is described by the operator $\frac{D}{\ell}$ as discussed above. As pointed out above the finite action of $\frac{D}{\ell}$ corresponds to a unitary transformation in the Fock space whose geometrical meaning is:
$$
e^{i\frac{D}{\ell}t}\:\hat{\phi}(V,\theta,\varphi)\:e^{-i\frac{D}{\ell}t}:= \hat{\phi}(e^{\frac{t}{\ell}}V,\theta,\varphi).
$$
Since the reference state $\ket{0}_\fH$ of theory is quasi free and the one-point function on $\ket{0}_\fH$ is zero,
expectation values of products of fields can be computed by combination of expectation values of products of couples of fields.
In this case the KMS condition needs to verified only for two point functions.
That is 
\beq\label{kms}
\langle \pa\hat{\phi}_\fH(V,\theta,\varphi) \alpha_{t+i\beta} ( \pa\hat{\phi}_\fH)(V',\theta',\varphi')  \rangle_\fH =
\langle \alpha_t (\pa\hat{\phi}_\fH)(V,\theta,\varphi)\pa\hat{\phi}_\fH(V',\theta',\varphi')  \rangle_\fH,
\eeq
where $\alpha_t(\pa\hat{\phi}_\fH)=e^{it\frac{D}{\ell}}\:\pa\hat{\phi}_\fH\:    e^{-it\frac{D}{\ell}}$.
For the discussion of section four $\alpha_t(\pa\hat{\phi}_\fH(V,\theta,\varphi))=\pa\hat{\phi}_\fH(Ve^{\frac{t}{\ell}},\theta,\varphi)$ for real $t$.
Proving the condition \nref{kms} is equivalent to study the analytic continuation (in $t$) of the two-point function $G_\fH(e^{\frac{t}{\ell}}V,V')$ \nref{twopoint}.
To this end notice that 
$$
G_\fH(Ve^{i2\pi},V')=G_\fH(V,V').
$$
Thus by direct computation, one finds that the condition \nref{kms} is verified for $\be=2\pi\ell$.
We conclude that vacuum $\ket{0}_\fH$ is thermal at inverse temperature $\be=2\pi\ell$ w.r. to $\frac{D}{\ell}$.
Since there is a unitary relation between with the dS quantum field theory, that thermality holds also for the quantum field theory in the dS space w.r. to the generator of boost. In fact $\frac{D}{\ell}$ for the bulk theory is nothing but the generator of boosts $i\frac{\pa}{\pa t}$.
We stress once again that our computation was made using $\pa\hat{\phi}_\fH$ because of the form-smearing necessary to write a theory in a measure independent way.

\se{Conclusions}
We have shown that dS free massive quantum field theory enjoys a holographic relation with quantum field theory defined on the dS cosmological horizon. This holographic relation corresponding to a filed restriction is also a unitary relation.
The dual horizon theory has a manifest conformal symmetry ($SL(2,\bR)$ invariance).
Moreover, using holography, the proof of the validity of KMS condition for dS quantum field theory becomes turned out simple.

\vspace{0.5cm}
\appendix
\noindent
{\bf \Large Appendix}
\se{de Sitter field in static coordinates}
We shown explicitly that the holographic relation holds choosing another reference state.
To this hand consider the general solution of the KG equation $\Box \psi = \la^2\psi$ \nref{KGplan} in static coordinate \nref{static}. 
It reads 
\beq\label{staticwave}
\psi(t,r,\theta,\phi)=\int_{\bR^+} e^{-iEt} \sum_{lm} \at\frac{r}{\ell}\ct^l \at 1-\frac{r^2}{\ell^2}\ct^{i\ell E/2}  
\frac{R_{El}^\nu (r)}{\sqrt{2E}}\;
 Y^l_m(\theta,\phi)\;\widetilde{\psi^+_{lm}}(E)\;dE  + \cc
\eeq
where 
$$
R_{El}^\nu(r)=A_{El} F\at \frac{l+\frac{3}{2}+i\ell E -\nu}{2}, \frac{l+\frac{3}{2}+i\ell E +\nu}{2} ;l+\frac{3}{2};\frac{r^2}{\ell^2} \ct 
$$
$A_{El}$ is a normalization constant, as before $\nu=\sqrt{9/4+\la^2\;\ell^2}$, encompasses information about particle mass and coupling curvature constant. $F$ is the hypergeometric function.
Notice that to solve the equation of motion we have imposed that the wave function vanish at the origin of the sphere 
(where the coordinate system is not well defined).
Our real wavefunctions are the real smooth function \nref{staticwave} satisfying the Klein Gordon equation in static coordinates.
The one-particle Hilbert space $\cH_>$ is defined as the completion of the positive frequency part of the wavefunctions \nref{staticwave} w.r. to 
the scalar product in static coordinate 
$\langle\psi_+,\psi'_+\rangle:=
i\int_\Sigma \at\: \overline{\psi_+}\: \nabla_\mu \psi'_+- {\psi'_+}\: \nabla_\mu \overline{\psi_+}\ct n^\mu \sqrt{\sigma} d\sigma 
$ where now $\Sigma$ is a $t-$constant hypersurface. $\Si$ is a Cauchy surface for the region $2$ (see the Figure).
In static coordinates the horizon is located at $r=\ell$. 
Form the one-particle Hilbert space the quantum field theory on the the bosonic Fock space $\gF_+(\cH_>)$ arises in a straightforward way.
We do not enter in that details, 
on the other hand we search for a holographic relation with a horizon quantum theory.
Here the horizon $\fH^>$ is only half of the cosmological horizon $\fH$, precisely the part of $\fH$ bounding the region 2 of the Figure.
We concentrate us on the behavior of the mode decomposition \nref{staticwave} 
near the cosmological horizon. It reads: $\sim e^{-iEt}/{\sqrt{4\pi E}} \at\frac{\ell+r}{\ell-r}\ct^{iE/2}\; e^{i\tilde{\rho}_l}$.
That suggests the presence of a unitary relation between that theory and an horizon theory defined in the outer part $\fH^>$ of the cosmological horizon $\fH$. A point of $\fH$ belongs to $\fH^>$ if its null coordinate $V$ satisfies $V>0$ hence a natural null coordinate on $\fH^>$ is
$v=\log{V}=t+\frac{1}{2}\log{\at {\ell+r}{\ell-r} \ct} $. Everything said in section 4 for $\fH$ using the coordinate $V$, can be translated for $\fH^>$ using the coordinate $v$ building up a bosonic quantum field theory on $\gF_+(\cH_{\fH^>})$. 
Moreover there is a unitary transformation mapping the static particle $\psi_+\in\cH_>$ (located in the region 2 of the Figure) to the null one:
$$
(U\psi_+)(t,r,\theta,\varphi):=\psi_+(v,\theta,\varphi)=\left.\psi_+(t,r,\theta,\varphi)\right|_{\fH^>}\;,
$$
as in planar coordinate this unitary transformation is a restriction. 
As expected, translating that unitary relation to the Fock spaces, the holographic relation holds also in that coordinate system (considering a different reference state $\ket{0}_{\fH^>}$).
Notice that this result is not a consequence of the holographic procedure discussed in section 5, in fact 
the Fock space $\gF_+(\cH_{\fH^>})$ equipped with its natural vacuum $\ket{0}_{\fH^>}$ is not unitarily equivalent to $\gF_+(\cH_\fH)$ with $\ket{0}$ restricted to the points whose null coordinate is $V>0$. That because $\ket{0}_{\fH^>}$ appears as a non normalizable state in $\gF_+(\cH_\fH)$.

\vspace{0.5cm}
\small 
\noindent {\bf Acknowledgments}. 
The author is grateful to V.~Moretti for stimulating discussions, interesting suggestions and for reading the manuscript.
This work has been funded by Provincia Autonoma di Trento by the project FQLA, Rif. 2003-S116-00047 Reg. delib. n. 3479 \& allegato B.

\end{document}